\documentclass[a4paper]{jpconf}
\usepackage{graphicx}

\begin{document}
\title{{\underline {\bf MIMAC}}: MIcro-tpc MAtrix  of Chambers for dark matter directional detection}

\author{D.~Santos,  G.~Bosson, J.L.~Bouly, O.~Bourrion, Ch.~Fourel,\\O.~Guillaudin, J.~Lamblin, F.~Mayet, J.F.~Muraz, J.P.~Richer, Q.~Riffard}
\address{  LPSC, Universite Joseph Fourier Grenoble 1, CNRS/IN2P3, Institut Polytechnique de Grenoble, 53, Av. des Martyrs, 38026 Grenoble, France}

\author{L.~Lebreton, D.~Maire}
\address{	LMDN, IRSN Cadarache, 13115 Saint-Paul-Lez-Durance, France} 
\author{J. Busto, J. Brunner, D. Fouchez}
\address{CPPM, Aix-Marseille Universit\'e, CNRS/IN2P3, Marseille, France}

\ead{Daniel.Santos@lpsc.in2p3.fr}

\begin{abstract}

Directional detection of non-baryonic Dark Matter is a promising search strategy for discriminating  WIMP events from neutrons, the ultimate background for dark matter direct detection. This strategy requires both a precise measurement of the energy down to a few keV and 3D reconstruction of tracks down to a few mm. 
The MIMAC (MIcro-tpc MAtrix of Chambers) collaboration has developed in the last years an original prototype detector based on the direct coupling of large pixelized micromegas with a special developed fast self-triggered electronics showing the feasibility of a new generation of directional detectors. The first bi-chamber prototype has been installed at Modane, underground laboratory in June 2012. The first undergournd background events,  the gain stability and calibration are shown. The first spectrum of nuclear recoils showing 3D tracks coming from the radon progeny is presented.
\end{abstract}

\section{Introduction}

Directional detection of Dark Matter is based on the fact that the solar system moves with respect to the center of our galaxy  \cite{spergel}. Taking into account the hypothesis of the existence of a galactic halo of DM formed by WIMPs (Weakly Interacting Particles), we can expect a privileged direction for the nuclear recoils in our detector, coming from elastic collision with those WIMPs \cite{Billard1, Billard2}. 

The MIMAC (MIcro-tpc MAtrix of Chambers) detector project \cite{MIMAC} tries to get these elusive events by a double detection: ionization and track, at low gas pressure with low mass target nuclei (H, $^{19}$F ). In order to have a significant cross section with such low mass nuclei, we explore the axial, spin dependent, interaction on odd nuclei. The very weak correlation between the neutralino-nucleon scalar cross section and the axial one, as it was shown in \cite{PLB} and more recently in \cite{DAlb} makes this research, at the same time, complementary to the massive target experiments.

\section{MIMAC bi-chamber prototype}

The MIMAC  bi-chamber prototype is composed of two chambers sharing the same cathode being the module of the matrix devoted  to show the ionization and track measurement performances needed to achieve the directional detection strategy.. The prototype active volume ($V  \sim 5.8\,\mathrm{l}$) is filled with: $\mathrm{ 70\,\%\, CF_4  + 28\,\%\, CHF_3 + 2\,\% \,C_4H_{10}}$  at a pressure of $50\,\mathrm{mbar}$.
The primary electron-ion pairs produced by a nuclear recoil in one chamber of the matrix are detected by driving the electrons to the grid of a bulk micromegas\cite{bulk} and producing the avalanche in a very thin gap (256 $\mu$m).

\begin{figure}[h!]
\begin{minipage}{0.65\linewidth}
\centerline{\includegraphics[width=\linewidth]{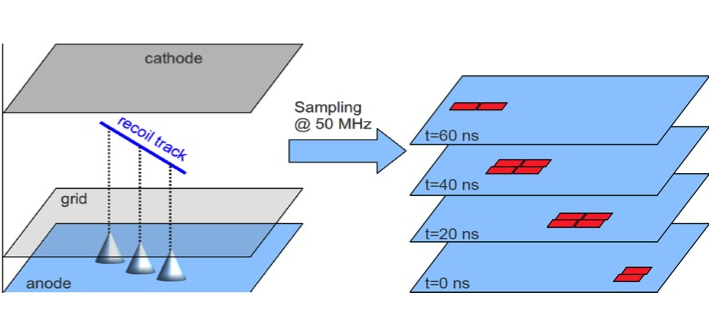}}
\end{minipage}
\hfill
\begin{minipage}{0.33\linewidth}
\centerline{\includegraphics[width=0.9\linewidth]{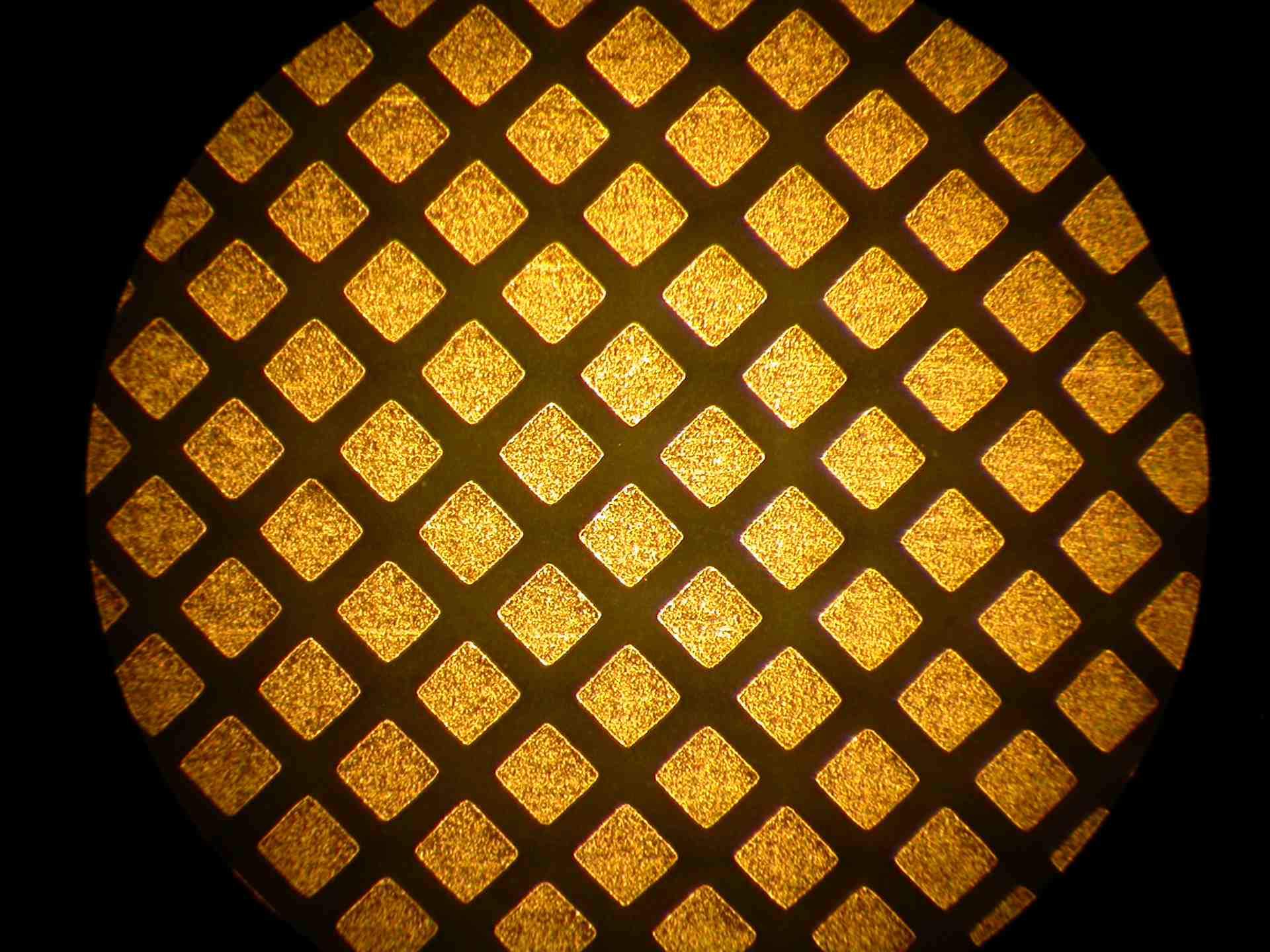}}
\end{minipage}
\caption{Left: Track reconstruction in MIMAC. The anode is read every 20 ns. The 3D track is reconstructed, from the consecutive number of images defining the event. Right: A picture of a small part of the pixellized micromegas designed by the IRFU (Saclay) team showing the 200 $\mu$m wide pixels \cite{Iguaz2011yc}.}
\label{fig:ShemaMicromegas}
\end{figure}

As pictured in figure  \ref{fig:ShemaMicromegas} (left), the primary electrons are collected to the grid in the drift space (25 cm)  and are multiplied by avalanche to the anode thus allowing to get 
information on X and Y coordinates.
To access to the X and Y coordinates with enough spatial resolution, a bulk micromegas was developped  \cite{Iguaz2011yc} with a 10.8 cm by 10.8 cm active area, segmented in pixels 200 $\mu$m wide, defining a 424 $\mu$m  pitch orthogonal structure shown in figure   \ref{fig:ShemaMicromegas} (right).
 In order to reconstruct the third dimension Z of a recoil track, the LPSC developed a self-triggered electronics able to perform the anode sampling at a frequency of 50 MHz.
This includes a dedicated 64 channels ASIC \cite{richer} associated to a DAQ \cite{bourrion} specially developped for the project.
 The pixellized micromegas coupled to the MIMAC electronics running the 512 channels, is in fact the validation of the feasibility of a large TPC for directional detection.

\section{MIMAC bi-chamber prototype at the Laboratoire Souterrain de Modane}

\begin{figure}[h]
\begin{minipage}{0.5\linewidth}
\includegraphics[scale= 1.15]{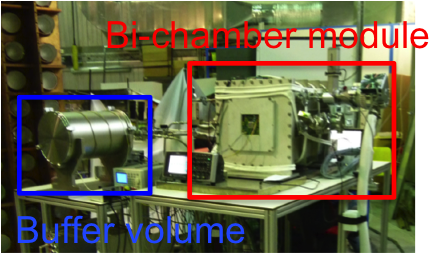}
\end{minipage}
\begin{minipage}{0.5\linewidth}
\includegraphics[scale= 0.4]{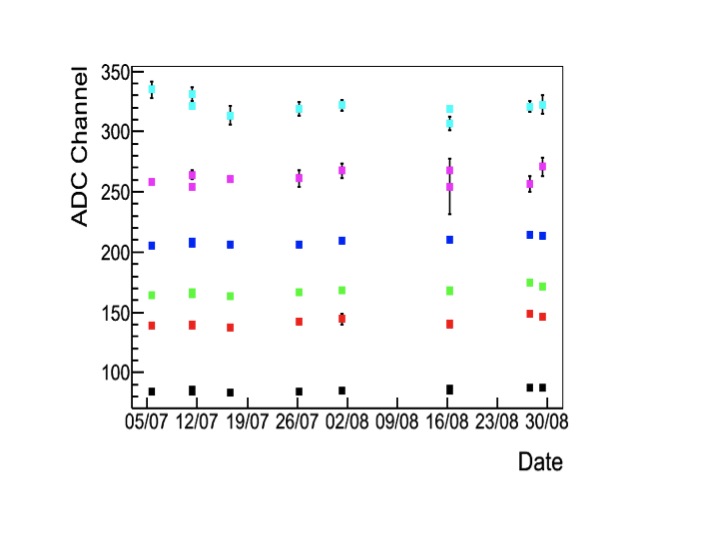}
\end{minipage}
\caption{Left: The bi-chamber prototype at the Laboratoire Souterrain de Modane in June 2012. The bi-chamber module is identified in red and the buffer volume in blue. Right: The position of peaks of  Cd (3.2 keV), Cr (5.4 keV), Fe (6.4 keV), Cu (8.1 keV) and Pb (10.5 and 12.6 keV) fitted by a linear calibration in ADC channels as a function of time, highlighting the gain stability during the data taking period.}
\label{fig:detect}
\end{figure}

The MIMAC bi-chamber prototype was installed at the Laboratoire Souterrain de Modane (LSM) in June 2012, see fig. \ref{fig:detect} (left).
By means of an X-ray generator, the detector is weekly calibrated by Cd (3.2 keV), Cr (5.4 keV), Fe (6.4 keV), Cu (8.1 keV) and Pb (10.5 and 12.6 keV)  fluorescence photons.
Fig.~\ref{fig:detect} right shows the position of peaks fitted by a linear calibration in ADC channels as a function of time, highlighting the gain stability during the data taking period. The last two peaks from  Pb X-rays suffer of an important statistical fluctuation due to the low cross section interaction at 50 mbar in our gas mixture. 
The gas circulation system including a buffer volume, an oxygen filter, a dry pump and a pressure regulator allows to keep the gas quality stable in a closed circuit. Indeed, the presence of impurities and $\mathrm{O_2}$ must be controlled to prevent gain and energy resolution degradation.

\section{Preliminary analysis of the first months of data taking}
\label{sec:Modane}

\begin{figure}[h]
\begin{minipage}{0.55\linewidth}
\includegraphics[scale =0.43]{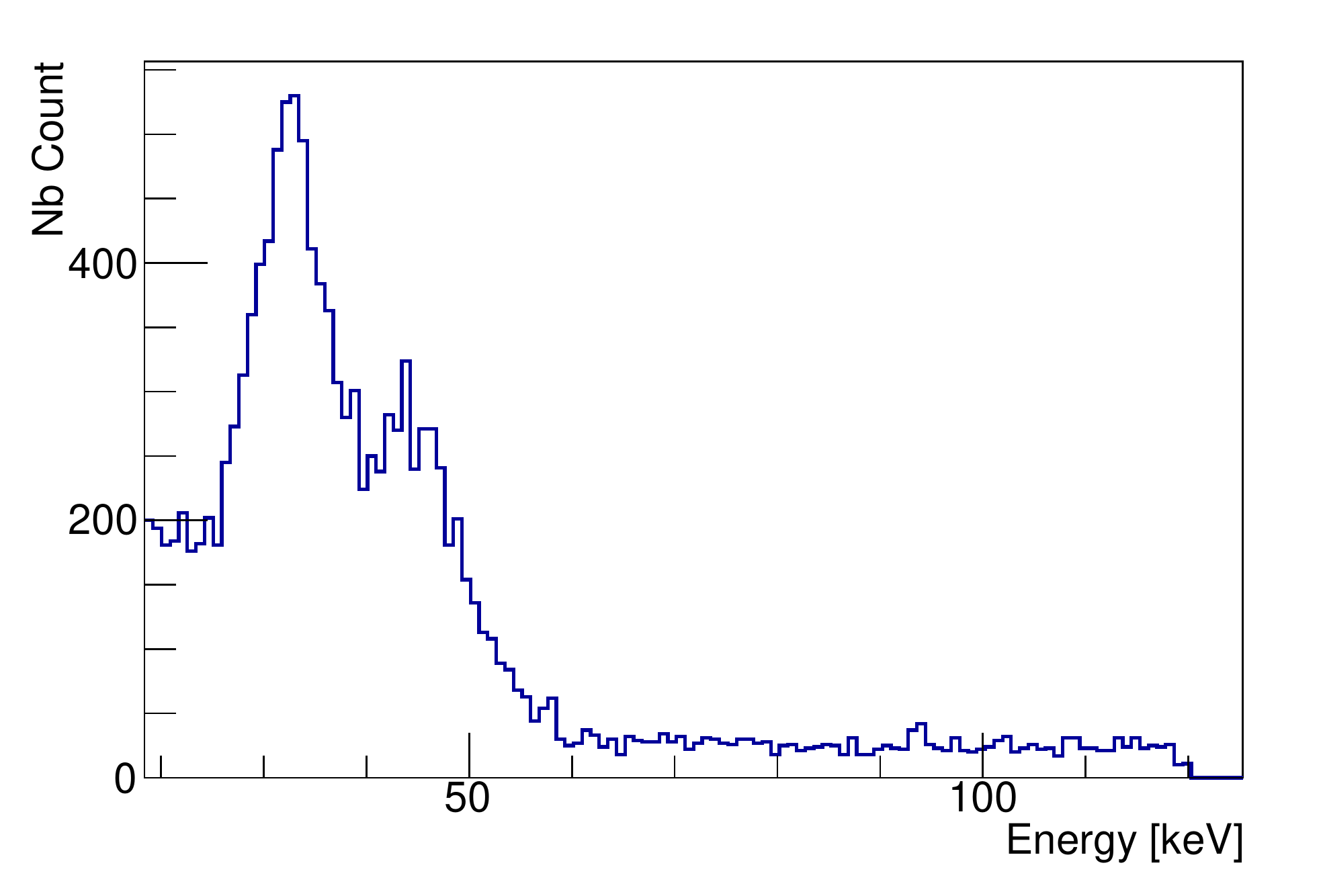}
\end{minipage}
\begin{minipage}{0.35\linewidth}
\includegraphics [scale =0.35]{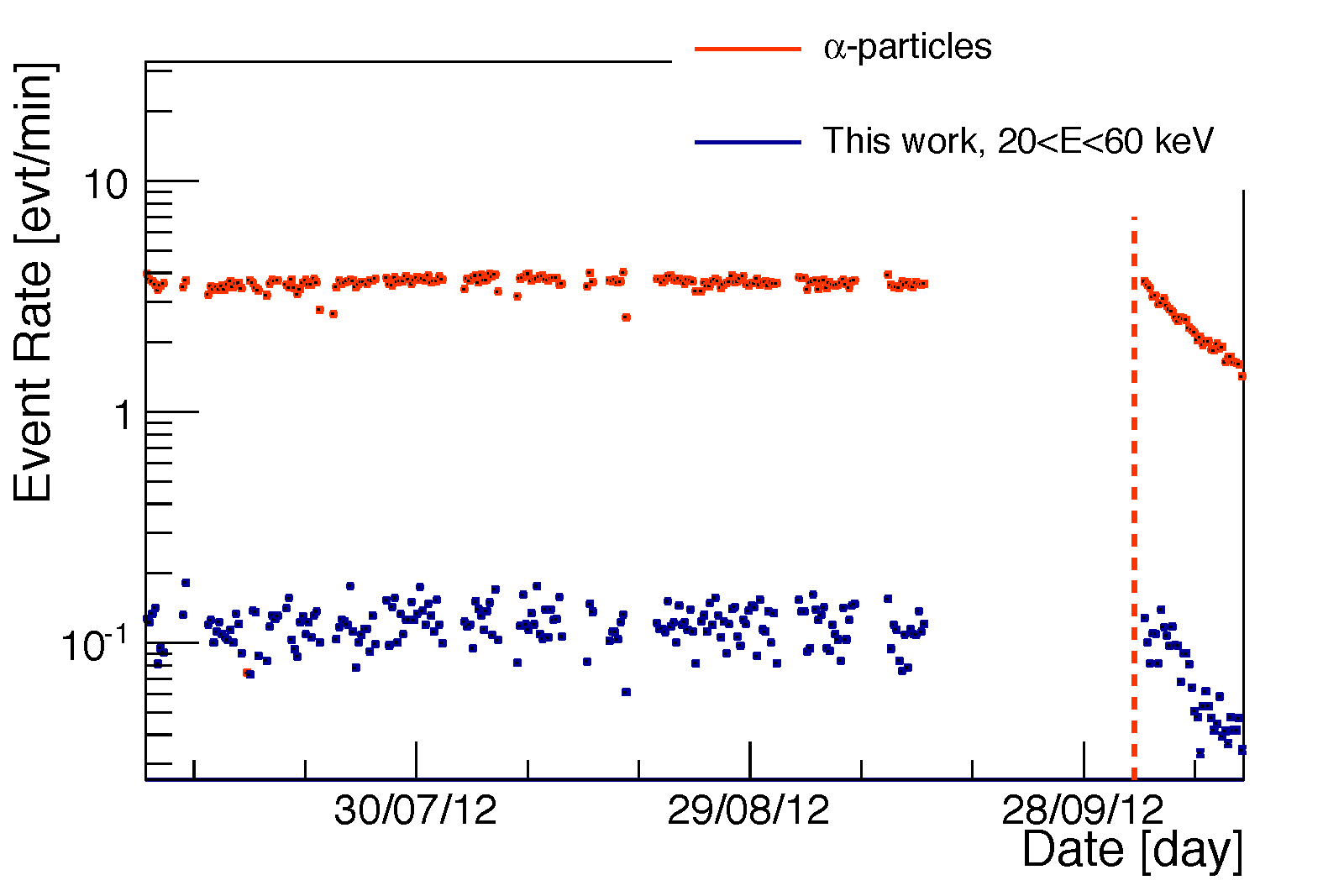}
\end{minipage}
\caption{ Left:  The energy spectrum obtained after a preliminary analysis between 20 and 120 keV.  Right:  The event rates of $\alpha$-particles (red dots) and the event rate of events selected by this analysis with energies between 20 and 60 keV (blue dots). The vertical dashed red line indicates the shutdown of the gas circulation system in October $\mathrm{3^{rd}}$, 2012. }
\label{fig:ModaneInstal}
\end{figure}

The first available data set of the bi-chamber prototype was started on July $\mathrm{5^{th}}$ 2012 and finished on October $\mathrm{12^{th}}$ 2012. 
The total  event rate measured with an ionization threshold of  2 keV and at least one (X ,Y) coincidence on the anode  was $5.6\pm 0.4\,\, \mathrm{evts/min}$. 
We performed a first data analysis 
that allowed us to obtain 3D-track events with an energy spectrum shown in Fig.~\ref{fig:ModaneInstal} (left). In this spectrum, we can clearly identify two peaks at $32.4\pm 0.1$ and $44\pm 0.3$ keV, and a flat distribution above 60 keV.

Fig.~\ref{fig:ModaneInstal} right shows the event rate of $\alpha$-particles (ADC saturation above 120 keV in red) and of events with energies between 20 and 60 keV selected by this analysis in blue.
These event rates remained constant until October $\mathrm{3^{rd}}$, when the gas circulation was stopped to isolate the bi-chamber from the gas circulation system. We can observe an exponential reduction of the event rates which fit with the $\mathrm{^{222}Rn}$ half-life $(T_{1/2} = 3.8\,\mathrm{days})$.   
It signs a pollution of the gas mixture by Radon isotopes contained in the circulation gas system due to a leak in the dry pump assuring the circulation of the gas from the buffer volume.
 The exponential decrease being dominated by the half-life of the  $\mathrm{^{222}Rn}$,  
 the contribution from the decay of $\mathrm{^{220}Rn}$ may be neglected in a first order approximation.

The seven $\alpha$-particles from the decay of $\mathrm{^{222}Rn}$ and $\mathrm{^{220}Rn}$ descendants, called Radons progeny, are emitted with kinetic energies $E^{kin}_{\alpha}$ ranging from 5.5 to 8.8 MeV.
A simulation showed that these $\alpha$-particles can pass through the 24 $\mu$m cathode to reach the other chamber even if the decay is produced in the gas volume. The simulated energy spectrum of the $\alpha$-particles reaching the other chamber was flat from 0.3 to 120~keV , our ionization energy dynamic range. 
In addition, if the decays occur at the  surfaces and if the $\alpha$-particles are absorbed in the matter, only the recoils of the daughter nuclei will be detected.
The recoils of the daughter nuclei from the Radon progeny have  kinetic energies from 100 to 170~keV, see \cite{Burgos} and \cite{QR}.

\begin{figure}[h]
\begin{center}
\includegraphics[scale=0.62]{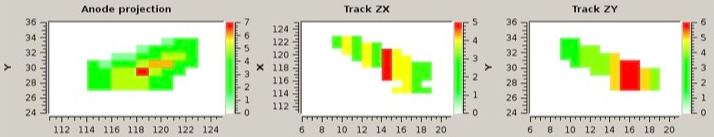}
\caption{A 3D recoil track measured at 34 keV in ionization: On the left, the X-Y plane of the anode showing the intersections of the strips fired. On the centre, the X projection as a function of time, every 20 ns. On the right, the same but for the Y projection}
\label{recoilevent}
\end{center}
\end{figure}

The difference between the kinetic energy of an ion $E^{kin}_{recoil}$ and the energy released by ionization $E^{ioni}_{recoil}$  in the active volume is determined by the Ionization Quenching Factor ({\it IQF})  defined as the ratio between the ionization energy released by a recoil and the ionization energy released by an electron at the same kinetic energy.
Taking into account the {\it IQF} correction from SRIM~\cite{SRIM}, the recoils from $\mathrm{^{222}Rn}$ and $\mathrm{^{220}Rn}$ progeny would be detected with ionization energies from 38 to 70 keV.
It was shown that SRIM  overestimates the {\it IQF} roughly by 20 \% at such recoil energies~\cite{Guillaudin:2011hu, santosQuenching}, in the case of $^4$He nuclei. There is no experimental data available for the quenching of  $^{218}$Po, $^{214}$Pb or $^{210}$Pb recoils in our gas mixture. This measurment may be one of the first evidence of the quenching in ionization for these heavy nuclei in a gas mixture. Taking into account the {\it IQF} estimated overestimation, the energies of the recoils from the $\mathrm{^{222}Rn}$ progeny  fit the spectrum. 
We can confidently conclude  that the 3D-track events associated to the energy spectrum shown in Fig.~\ref{fig:ModaneInstal} (left) come from the $\mathrm{^{222}Rn}$ progeny in the detector. 
The  identification of the contribution of each nuclear recoil from the $\mathrm{^{222}Rn}$ progeny will be presented in a future paper.
As an illustration of the high quality of data obtained at Modane, we show in fig.\ref{recoilevent} a very interesting recoil event of 34 keV in ionization from the radon progeny discussed above.

For directional detection it is important to discriminate Radon progeny events from WIMP like events. This discrimination should be possible by using the time and spatial coincidence between the two chambers. For those events produced at the cathode surface, we expect one recoil event in one of the chambers and the $\alpha$-particle through the cathode in the other one.
For this data taking period, the time synchronization of the chambers was not available.
  The detector was upgraded and the new version was installed end of June 2013. The upgraded version has a three times bigger gas buffer volume, a thinner cathode (12  $\mu$m  instead of 24  $\mu$m), a charcoal filter at -15 $^o$C for radon filtering and the electronic synchronisation of the chambers. The evalutation of the rejection of the nuclear recoil events from the radon progeny by the coincidences between the two chambers will be performed in the next data taking period started in July 1$^{st}$ 2013.

\begin{figure}[h]
\begin{center}
\includegraphics[scale=0.9]{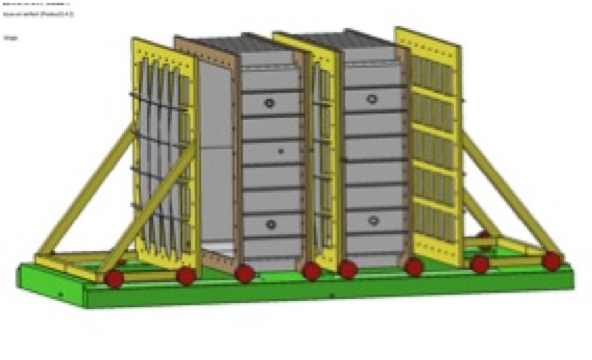}
\caption{The preliminary mechanical design of the demonstrator of  MIMAC -$1\,\mathrm{m^3}$ .}
\label{1m3}
\end{center}
\end{figure}

The next step in the definition of a large TPC for directional detection is to build a demonstrator of  $1\,\mathrm{m^3}$. A preliminary mechanical design keeping the  bi-chamber module as the elementary structure repeated with larger micromegas detectors (20 cm x 20 cm)  is shown in fig.\ref{1m3}.

\section{Conclusions}

The MIMAC detector provides the ionization energy of a recoiling nucleus at  low energies and the reconstruction of its 3D track. 
The  bi-chamber prototype was installed at the LSM in June 2012.
A preliminary analysis of the first data set allowed us to observe, for the first time, the 3D-track and energy of recoils from the $\mathrm{^{222}Rn}$ progeny.
The next step of the MIMAC project  will be the development of the MIMAC $1\,\mathrm{m^3}$. The 1 $\mathrm{m^3}$ detector will be the demonstrator for a large TPC devoted to DM directional search.



\end{document}